\documentclass[letterpaper]{JHEP3}
\usepackage{graphicx}

\title{Top Hypercharge}

\author{Cheng-Wei Chiang \\
  Department of Physics and Center for Mathematics and Theoretical
  Physics, National Central University, Chungli, Taiwan
  320, R.O.C. and \\
  Institute of Physics, Academia Sinica, Taipei, Taiwan 115, R.O.C.}

\author{Jing Jiang \\
  Institute of Theoretical Science, University of Oregon, Eugene, OR
  97403, U.S.A.}

\author{Tianjun Li \\
  George P. and Cynthia W. Mitchell Institute for Fundamental Physics,
  Texas A$\&$M University, College Station, TX
  77843, U.S.A. and \\
  Institute of Theoretical Physics, Chinese Academy of Sciences,
  Beijing 100080, P. R. China}

\author{Yong-Rui Wang \\
  Institute of Theoretical Physics, Chinese Academy of Sciences,
  Beijing 100080, P. R. China}

\date{\today}

\abstract{We propose a top hypercharge model with gauge symmetry
  $SU(3)_C \times SU(2)_L \times U(1)_1 \times U(1)_2$ where the first
  two families of the Standard Model (SM) fermions are charged under
  $U(1)_1$ while the third family is charged under $U(1)_2$.  The
  $U(1)_1 \times U(1)_2$ gauge symmetry is broken down to the $U(1)_Y$
  gauge symmetry, when a SM singlet Higgs field acquires a vacuum
  expectation value.  We consider the electroweak constraints, and
  compare the fit to experimental observables to that of the SM.  We
  study the quark CKM mixing between the first two families and the
  third family, the neutrino masses and mixing, the flavour changing
  neutral current effects in meson mixing and decays, the $Z'$
  discovery potential at the Large Hadron Collider, the dark matter
  with a gauged $Z_2$ symmetry, and the Higgs boson masses.}

\preprint{MIFP-07-25}


\begin{document}

\section{Introduction \label{sec:intro}}

There are many motivations for physics beyond the Standard Model (SM).
For example, the fine-tuning problem such as the gauge hierarchy
problem leads to supersymmetry~\cite{Dimopoulos:1981au},
technicolor~\cite{Weinberg:1975gm}, extra
dimensions~\cite{ArkaniHamed:1998rs, Randall:1999ee}, and other
models.  Aesthetic considerations such as the unification of the
fundamental interactions and the explanation of the charge
quantization lead to Grand Unified Theories (GUTs)~\cite{Pati:1974yy}
and string theory~\cite{GSW, JP}.  In this paper, we choose to neglect
the fine-tuning and aesthetic concerns, and concentrate on the
electroweak precision measurements from a phenomenological point of
view.

As we learn from the particle data book, there are four measurements
with significant deviations: $\sigma_{\rm had}$, $A_{FB}^{(0,b)}$,
$A_e$ and $g_L^2$ have $2.0$ $\sigma$, $-2.4$ $\sigma$, $2.0$
$\sigma$, and $-2.7$ $\sigma$ deviations (pull from experimental
values), respectively~\cite{Yao:2006px}. A simple solution to such
deviations is a $U(1)'$ model where an additional $U(1)'$ gauge
symmetry is introduced~\cite{Erler:1999nx}.  Moreover, the masses of
the third family fermions are relatively larger than those of the
first two families.  The quark CKM mixings between the first two
families and the third family are small while the neutrino mixings are
bilarge. These facts may imply that the first two families and the
third family may have different $U(1)'$ charges, and the right-handed
neutrinos might be neutral under $U(1)'$ so that the seesaw mechanism
still works~\cite{seesaw}.  To avoid the introduction of a global
symmetry which can be broken via quantum gravity effects, the model
may contain a gauged $Z_2$ symmetry, and a field charged odd under the
symmetry can be a dark matter candidate.  Interestingly, the
additional $U(1)'$ symmetry can easily generate the required gauged
$Z_2$ symmetry after it is broken.  In general, the $U(1)'$ model is
well motivated from the superstring constructions~\cite{string},
four-dimensional GUTs~\cite{review}, higher dimensional orbifold
GUTs~\cite{Orbifold}, as well as models with dynamical symmetry
breaking~\cite{Hill:2002ap}.

In this paper, we propose a top hypercharge model where the first two
families of the SM fermions are charged under $U(1)_1$ and the third
family is charged under $U(1)_2$.  We note in passing that top color
and top flavour models have been considered in Refs.~\cite{Topcolor}
and~\cite{Topflavor}, respectively.  The $U(1)_1\times U(1)_2$ gauge
symmetry is equivalent to the $U(1)_Y\times U(1)'$ gauge symmetry
after an $SO(2)$ rotation of the $U(1)_1\times U(1)_2$ gauge
fields. The $U(1)_1\times U(1)_2$ gauge symmetry is broken down to the
$U(1)_Y$ by giving a vacuum expectation value (VEV) to a SM singlet
Higgs field. The $SU(2)_L\times U(1)_Y$ gauge symmetry is further
broken down to the electromagnetic $U(1)_{EM}$ gauge symmetry by
giving VEV's to SM doublets.  We calculate the neutral gauge boson
masses and their corresponding mass eigenstates, and the charged gauge
boson masses.  Moreover, we consider electroweak constraints, and
perform a $\chi^2$ analysis for various experimental observables.  Our
model achieves a better fit than the SM.  To generate the quark CKM
mixings between the first two families and the third family, we need
to introduce an additional SM Higgs doublet or extra vector-like
particles.  We consider the flavour-changing neutral current (FCNC)
effects in $B$ physics, and obtain the corresponding constraints.  In
addition, we consider the $Z'$ production and its discovery potential
at the Large Hadron Collider (LHC), a dark matter candidate with the
gauged $Z_2$ symmetry, and the Higgs boson masses.

This paper is organized as follows. We present the model in Section
\ref{sec:model}, consider the electroweak constraints in Section
\ref{sec:EW}, and study the quark CKM mixings and FCNC effects in
Section \ref{sec:FCNC}.  The $Z'$ production, dark matter and the
Higgs boson masses are discussed in Sections \ref{sec:prod} and
\ref{sec:DM-Higgs}, respectively. Our conclusions are given in Section
\ref{sec:summary}.

\section{The model \label{sec:model}}

The top hypercharge model is based on the gauge group $SU(3)_C\times
SU(2)_L \times U(1)_1 \times U(1)_2$. The first two families of the SM
fermions are charged under $U(1)_1$ while the third family is charged
under $U(1)_2$.  The $U(1)_1\times U(1)_2$ gauge symmetry is
equivalent to the $U(1)_Y\times U(1)'$ gauge symmetry after an $SO(2)$
rotation of the gauge fields. The representations of the SM fermions
under the $SU(3)_C\times SU(2)_L \times U(1)_1 \times U(1)_2$ gauge
symmetry are as follows
\begin{eqnarray}
& Q_{i L}: (3, 2,~1/6,~0) ~, \hspace{1cm} Q_{3 L}: (3, 2,~0,~1/6) ~, \nonumber \\
 & u_{i R}: (3, 1,~2/3,~0) ~,\hspace{1cm} u_{3 R}: (3, 1,~0,~2/3) ~, \nonumber \\
 & d_{i R}: (3, 1,~-1/3,~0) ~,\hspace{1cm} d_{3 R}: (3, 1,~0,~-1/3) ~, \nonumber \\
 & L_{i L}: (1, 2,~-1/2,~0) ~, \hspace{1cm} L_{3 L}: (1, 2,~0,~-1/2)~,  \nonumber \\
 & e_{i R}: (1, 1,~-1,~0) ~,\hspace{1cm} e_{3 R}: (1, 1,~0,~-1) ~, \nonumber \\
 & N_k : (1, 1,~0,~0) ~,~\,
\end{eqnarray}
where $i=1,2$, and $k=1,2,3$.  $Q$ and $L$ denote the left-handed
$SU(2)_L$ doublets of quarks and leptons, respectively.  $u$, $d$,
$N$, and $e$ denote the right-handed up-type quark, down-type quark,
right-handed neutrino, and charged lepton, respectively. The
right-handed neutrinos $N_k$ with intermediate-scale masses are
included to account for the neutrino masses and mixings via the seesaw
mechanism~\cite{seesaw}. With the right-handed neutrinos, it is
possible to generate baryon asymmetry via
leptogenesis~\cite{Fukugita:1986hr}.

The covariant derivative is written as
\begin{eqnarray}
D_\mu &=& \partial_\mu - i g A^a_\mu T^a - i g^{\prime}_1 B^1_\mu
Y_1 - i g^{\prime}_2 B^2_\mu Y_2 \ ,
\end{eqnarray}
where $T^a$ are the $SU(2)_L$ generators, and $Y_{1(2)}$ is the
$U(1)_{1(2)}$ charge of the corresponding particle. In addition, $A^a$
and $B^{1(2)}$ are respectively the gauge bosons for $SU(2)_L$ and
$U(1)_{1(2)}$ gauge symmetries with the coupling constants $g$ and
$g^{\prime}_{1(2)}$.

The $SU(3)_C\times SU(2)_L \times U(1)_1 \times U(1)_2$ symmetry needs
to be broken down to the SM gauge symmetry.  It is accomplished by the
VEV of a complex scalar field $\Sigma$ , which transforms as $(1,
1,1/2,-1/2)$. The relevant term in the Lagrangian is $(D_\mu
\Sigma)^{\dagger}(D^\mu \Sigma)$. We set $\langle \Sigma
\rangle=u/\sqrt{2}$, then the gauge fields $B^1_\mu$ and $B^2_\mu$
acquire a mass-squared matrix
\begin{eqnarray}
\frac{u^2}{4}  \left( \begin{array}{cc}
                                   {g^\prime_1}^2 & -g^\prime_1 g^\prime_2  \\
                                   -g^\prime_1 g^\prime_2  &  {g^\prime_2}^2 \\
                                 \end{array}
                          \right)~~.
\end{eqnarray}
At the scale $u$, this renders a massless and a massive gauge bosons
\begin{eqnarray}
 B_\mu &=& \cos \phi B^1_\mu + \sin \phi B^2_\mu ~,
 \nonumber \\
\widetilde{B}_\mu &=& - \sin \phi B^1_\mu + \cos \phi B^2_\mu ~,
\label{B_definition}
\end{eqnarray}
where the mixing angle $\phi$ is defined by $\tan\phi = g'_1 / g'_2$,
with the corresponding masses
\begin{eqnarray}
m^2_{B_\mu} = 0 ~, \hspace{1cm} m^2_{\widetilde B_\mu} = {1 \over
4}({g^\prime_1}^2+{g^\prime_2}^2) u^2 ~.
\end{eqnarray}
The massless field $B_\mu$ corresponds to the hypercharge 
gauge field in the SM.

Using the inverse transformation, we write the covariant derivative in
terms of $B_\mu$ and $\widetilde{B}_\mu$ as
\begin{eqnarray}
D_\mu = && \partial_\mu - i g A^a_\mu T^a - i g' (Y_1+Y_2) B_\mu 
%\nonumber \\
%&& 
- i g' (-Y_1 \tan \phi + Y_2 \cot \phi) \widetilde{B}_\mu ~,
\label{co-derivative1}
\end{eqnarray}
where
\begin{eqnarray}
Y=Y_1+Y_2  ~,~~ 
\frac{1}{g^{\prime 2}} = \frac{1}{g_1^{\prime 2}} + \frac{1}{g_2^{\prime 2}}
~.~\,
\end{eqnarray}
In terms of the electron charge $e$, the analog of the weak mixing
angle $\theta$ in the SM, and the $SO(2)$ rotation angle $\phi$,
we can express the three coupling constants in the model as
\begin{eqnarray}
g=\frac{e}{\sin \theta} ~, \hspace{0.5cm} g^\prime_1=\frac{e}{\cos
\theta \cos \phi} ~,  \hspace{0.5cm} g^\prime_2=\frac{e}{\cos \theta
\sin \phi} ~.
\label{parameter}
\end{eqnarray}

Now we introduce two scalar Higgs doublets $\Phi_1$ and $\Phi_2$,
which transform under $SU(3)_C\times SU(2)_L \times U(1)_1 \times
U(1)_2$ as $(1, 2,1/2,0)$ and $(1, 2,0,1/2)$. When they acquire the
following VEVs
\begin{eqnarray}
\langle \Phi_1 \rangle = {1 \over \sqrt{2}} \left( \begin{array}{c}
                                 0 \\
                                 v_1
                               \end{array}
                         \right)~,~~~~~
\langle \Phi_2 \rangle = {1 \over \sqrt{2}} \left( \begin{array}{c}
                                 0 \\
                                 v_2
                               \end{array}
                         \right)~,
\end{eqnarray}
the $SU(2)_L\times U(1)_Y$ gauge symmetry is broken down to the
$U(1)_{EM}$.  $\Phi_1$ is responsible for the masses of the fermions
of the first two generations, and $\Phi_2$ the third generation. As
the hierarchy in the fermion mass spectrum indicates, one expects that
$v_1$ may be smaller than $v_2$. And we define
\begin{eqnarray}
\tan\beta \equiv {{v_2}\over {v_1}}~.~\,
\end{eqnarray}

The terms $(D_\mu \Phi_1)^{\dagger}(D^\mu \Phi_1)+(D_\mu
\Phi_2)^{\dagger}(D^\mu \Phi_2)$ in the Lagrangian lead to the
following mass terms for the $SU(2)_L\times U(1)_1\times U(1)_2$ gauge
bosons after $\Phi_1$ and $\Phi_2$ get VEVs
\begin{eqnarray}
\Delta {\cal L} &=& {1 \over 4} g^2 \left(v_1^2+v_2^2\right) W^+_\mu
W^{-\mu} \nonumber \\
 &+& {1 \over 8} \left( \begin{array}{ccc}
B_\mu^1 & B_\mu^2 & A_\mu^3 \end{array} \right) \left(
\begin{array}{ccc} {g_1^\prime}^2 v_1^2 & 0 & -g g_1^\prime v_1^2 \\
                 0 & {g_2^\prime}^2 v_2^2 & -g {g_2^\prime}^2 v_2^2
                 \\
                 -g g_1^\prime v_1^2 & -g g_2^\prime v_2^2 & g^2
                 (v_1^2 + v_2^2)
                 \end{array}
                 \right)
\left( \begin{array}{c} B^{1 \mu} \\
                    B^{2 \mu} \\
                    A^{3 \mu}
                    \end{array}
 \right) ~.
\end{eqnarray}
Therefore, the $W$ bosons get mass $m^2_W = {1 \over 4} g^2
(v_1^2+v_2^2)$, and the mass-squared matrix of the neutral gauge
sector becomes
\begin{eqnarray}
M_{neutral}^2 = \frac{u^2}{4} \left( \begin{array}{ccc}
{g^\prime_1}^2 (1+\epsilon_1) & -g_1^\prime g_2^\prime & -g g_1^\prime \epsilon_1 \\
  -g_1^\prime g_2^\prime  &  {g^\prime_2}^2 (1 + \epsilon_2) & -g g^\prime_2 \epsilon_2\\
 -g g_1^\prime \epsilon_1 & -g g_2^\prime \epsilon_2 & g^2 (
 \epsilon_1+ \epsilon_2)
\end{array}
\right) ~, \label{M_neutral}
\end{eqnarray}
where $\epsilon_1\equiv v_1^2/u^2$ and $\epsilon_2\equiv v_2^2/u^2$,
and the basis is $(B^1_\mu,~ B^2_\mu,~ A^3_\mu)$.  The matrix can be
diagonalized by an orthogonal matrix $R$
\begin{eqnarray}
R^T M_{neutral}^2 R = {\rm diag} \{0,~ m_Z^2,~ m_{Z^\prime}^2 \} ~,
\end{eqnarray}
where 0, $m_Z^2$ and $m_{Z^\prime}^2$ are the eigenvalues of
$M_{neutral}^2$. The 0 eigenvalue corresponds to the massless photon
field $A_\mu$, and the other two non-zero eigenvalues correspond to
$Z$ and $Z'$ gauge bosons. We use $Z$ to denote the ``light $Z$
boson'', which is the observed one, and $Z^\prime$ to denote the
``heavy $Z$ boson''.  The transformation between the weak eigenstates
and mass eigenstates is
\begin{eqnarray}
\left( \begin{array}{c} B_\mu^1 \\ B_\mu^2 \\ A_\mu^3 \end{array}
\right) = R \left( \begin{array}{c} A_\mu \\ Z_\mu \\ Z_\mu^\prime
\end{array} \right) ~.
\end{eqnarray}

To obtain the interactions between the SM fermions and gauge bosons,
we write the covariant derivative in terms of the mass eigenstates of
gauge bosons
\begin{eqnarray}
D_\mu &=& \partial_\mu - i {g \over \sqrt 2} \left(T^+
W^+_\mu + T^- W^-_\mu \right)
\nonumber \\
&-& i \left( g T_3 R_{32} + g_1^\prime Y_1 R_{12} + g_2^\prime Y_2
R_{22} \right) Z_\mu \nonumber \\
&-& i \left( g T_3 R_{33} + g_1^\prime Y_1 R_{13} 
+ g_2^\prime Y_2 R_{23} \right) Z^\prime_\mu \nonumber \\
&-& i e Q A_\mu~,
\label{co_derivative_mass}
\end{eqnarray}
where $T^\pm=(T^1 \pm i T^2)$, and $Q=T_3+Y=T_3+Y_1+Y_2$ is the
electric charge for the corresponding particle.

The Lagrangian of the SM fermion Yukawa couplings is
\begin{eqnarray}
 - {\cal L}_{Yukawa} &=& Y^u_{i} \bar {u}_{iR} \Phi_1 Q_{iL}
  + Y^u_{3} \bar {u}_{3R} \Phi_2 Q_{3L}
  + Y^d_{ij} \bar {d}_{iR} \widetilde \Phi_1 Q_{jL}
 + Y^d_{33} \bar {d}_{3R} \widetilde \Phi_2 Q_{3L}   \nonumber \\
 &+& Y^e_{i} \bar {e}_{iR} \widetilde \Phi_1 L_{iL}
 + Y^e_{3} \bar {e}_{3R} \widetilde \Phi_2
 L_{3L} + Y^\nu_{ki} N_k \Phi_1 L_i + Y^\nu_{k3} N_k \Phi_2 {L}_{3}
   \nonumber \\
 &+& M_{kl}^N N_k N_l+  h.c. ~,
\end{eqnarray}
where $i,j=1,2$, $k, l=1,2,3$, and $\widetilde \Phi = i \sigma_2
\Phi^*$.  In the presence of $\Phi_1$ and $\Phi_2$, there will be
Yukawa mixings between the first two generations only, for both quarks
and leptons.  Interestingly, there is no problem for the neutrino
mixings.  The point is that after the seesaw mechanism~\cite{seesaw},
the bilarge neutrino mixings can be mainly generated from the
neutrino sector via the mixings in the right-handed Majorana neutrino
mass matrix $M_{kl}^N$. In fact, after the seesaw mechanism, we obtain
the following dimension-5 operators
\begin{equation}
-{\cal L} = {{\Phi_1 L_i \Phi_1 L_j}\over {\Lambda_{ij}}}
+ {{\Phi_1 L_i \Phi_2 L_3}\over {\Lambda_{i3}}}
 + {{\Phi_2 L_3 \Phi_2 L_3}\over {\Lambda_{33}}}~,~\,
\end{equation}
where $\Lambda_{ij}$, $\Lambda_{i3}$, and $\Lambda_{33}$ are
intermediate mass scales related to the right-handed neutrino
masses. Moreover, the SM quark CKM mixings can be generated by
introducing extra Higgs doublets or heavy vector-like fermions, which
will be discussed in Section \ref{sec:FCNC}.

\section{Electroweak constraints \label{sec:EW}}

The model contains a new gauge boson $Z^\prime$, the mass of which can
be constrained by the electroweak observables.  We calculate in our
model the predicted values of the observables in the $\overline{MS}$
scheme, as used in Ref.~\cite{Yao:2006px}.  For all observables, we
compute the deviations from the SM at the tree level and then scale them
by the same loop corrections as in the SM, which are given in Table
10.1 of Ref.~\cite{Yao:2006px}.  The value of $\sin^2\theta$ in the
$\overline{MS}$ scheme is $\approx 0.2312$ for the SM.  In our model,
it receives a correction in order to reproduce the correct $Z$
mass.  Other inputs used here are $\hat\alpha(M_Z)^{-1}=127.904$ and
$v=246.3$ GeV.

In the model, the first two generations of fermions couple to the
regular $Z$ boson differently from the third one.  Therefore, if there
is mixing between these generations, there can be FCNC at the tree
level.  However, this is not seen to happen since the branching ratios
of the corresponding $Z$ decay modes are very small.  For example, we
have the experimental bounds \cite{Yao:2006px} $BR(Z \to e^\pm\mu^\mp)
< 1.7\times10^{-6}$, $BR(Z \to e^\pm\tau^\mp) < 9.8\times10^{-6}$ and
$BR(Z \to \mu^\pm\tau^\mp) < 1.2\times10^{-5}$ at 95\% confidence
level (CL).  There is also no evidence that the $Z$ boson decays to
quark pairs of different flavors.  Therefore, in the calculation of
$Z$ pole observables, we avoid considering $Z$-mediated FCNC.

The electroweak precision bounds can be consistent with our model
predictions as long as $\epsilon_1$ and $\epsilon_2$ are taken to be
very small. However, if $\epsilon_1$ and $\epsilon_2$ are very small,
the mass of $Z^\prime$ will be so large that it is beyond the LHC
reach in the near future.  Therefore, we will concentrate on the
parameter space where $M_{Z^\prime}$ is in the a few TeV range.  In
this section, for completeness, we will perform the analysis by
choosing $\tan\beta\equiv v_2/v_1$ equal to 0.5, 1, 2, 5, 10, 20, 35
and 50.  We emphasize that if $\tan\beta$ is small than 1, there
exists the Landau pole problem for the top quark Yukawa coupling below
the grand unification scale, string scale or Planck scale.  However,
the Landau pole problem can be solved if the cut-off scale or the
fundamental scale is low, for example, the large extra
dimensions~\cite{ArkaniHamed:1998rs}.  In addition, it is more natural
to take $\tan\beta > 1$, hence, we will concentrate on it when we
study the phenomenological consequences of our model in the following
sections.

\TABLE{
\caption{The experimental~\cite{Yao:2006px} and the predicted values
  of the $Z$-pole observables for the SM~\cite{Yao:2006px} and our
  model with $\tan\beta$=2 and 50 as two examples.  
  For best fits, the $\tan\beta=2$ case has $\cos^2\phi=4.3$ and
  $M_{Z^\prime}=2.4$ TeV, and the $\tan\beta=50$ case has
  $\cos^2\phi=1.22$ and $M_{Z^\prime}=10$ TeV.}
\vspace*{0.2cm}
\begin{tabular}{|c|c|c|c|c|c|c|c|}  \hline
Observables& Experimental data & \multicolumn{2}{|c|}{\mbox {SM}}
&\multicolumn{2}{|c|} {\mbox {$\tan\beta=2$}} &\multicolumn{2}{|c|} {\mbox {$\tan\beta=50$}}\\
\hline
     &    & best fit& pull & best fit & pull  & best fit & pull \\ \hline
     $M_W ({\rm GeV})$      & $80.450\pm0.058$ & $80.376$ & $1.3$ & $80.376$ & $1.3$ & $80.376$ & $1.3$\\
     $\Gamma_Z ({\rm TeV})$ & $2.4952\pm0.0023$ & $2.4968$ & $-0.7$ & $2.4972$ & $-0.9$ & $2.4971$ & $-0.8$\\
     $\sigma_{\rm had} [nb]$  & $41.541\pm0.037$ & $41.467$ & $2.0$ & $41.477$ & $1.7$ & $41.470$ & $1.9$ \\
     $R_e$           & $20.804\pm0.050$ & $20.756$ & $1.0$ & $20.7498$ & $1.1$ & $20.7534$ & $1.0$ \\
     $R_\mu$         & $20.785\pm0.033$ & $20.756$ & $0.9$ & $20.7498$ & $1.1$ & $20.7534$ & $1.0$ \\
     $R_\tau$        & $20.764\pm0.045$ & $20.801$ & $-0.8$ & $20.8110$ & $-1.0$ & $20.8035$ & $-0.9$ \\
     $R_b$           & $0.21629\pm0.00066$ & $0.21578$ & $0.8$ & $0.21570$ & $0.9$ & $0.21576$ & $0.8$ \\
     $R_c$           & $0.1721\pm0.0030$ & $0.17230$&$-0.1$ & $0.17231$ & $-0.1$ & $0.17230$ & $-0.1$ \\
     $A_e$           & $0.15138\pm0.00216$ & $0.1471$&$2.0$ & $0.1454$ & $2.8$ & $0.1463$ & $2.4$ \\
     $A_\mu$         & $0.142\pm0.015$ & $0.1471$&$-0.3$ & $0.1454$ & $-0.2$ & $0.1463$ & $-0.3$ \\
     $A_\tau$        & $0.136\pm0.015$ & $0.1471$ &$-0.7$ & $0.1484$ & $-0.8$ & $0.1472$ & $-0.7$ \\
     $A_b$           & $0.923\pm0.020$ & $0.9347$&$-0.6$ & $0.9348$ & $-0.6$ & $0.9347$ & $-0.6$ \\
     $A_c$           & $0.670\pm0.027$ & $0.6678$&$0.1$ & $0.6670$ & $0.1$ & $0.6674$ & $0.1$ \\
     $A_s$           & $0.895\pm0.091$ & $0.9356$&$-0.4$ & $0.9355$ & $-0.4$ & $0.9355$ & $-0.4$ \\
     $A^e_{FB}$      & $0.0145\pm0.0025$ & $0.01622$& $-0.7$ & $0.01584$ & $-0.5$ & $0.01604$ & $-0.6$ \\
     $A^\mu_{FB}$    & $0.0169\pm0.0013$ & $0.01622$ & $0.5$ &  $0.01584$ & $0.8$ & $0.01604$ & $0.7$ \\
     $A^\tau_{FB}$   & $0.0188\pm0.0017$ & $0.01622$& $1.5$ &  $0.01617$ & $1.5$ & $0.01614$ & $1.6$ \\
     $A^b_{FB}$      & $0.0992\pm0.0016$ & $0.1031$ & $-2.4$ & $0.1019$ & $-1.7$ & $0.1025$ & $-2.1$ \\
     $A^c_{FB}$      & $0.0707\pm0.0035$ & $0.0737$ & $-0.8$ & $0.0728$ & $-0.6$ & $0.0732$ & $-0.7$ \\
     $A^s_{FB}$      & $0.0976\pm0.0114$ & $0.1032$ & $-0.5$ & $0.1020$ & $-0.4$ & $0.1026$ & $-0.4$ \\
     $g_L^2$       & $0.30005\pm0.00137$ & $0.30378$ & $-2.7$ & $0.30398$ & $-2.9$ & $0.30390$ & $-2.8$ \\
     $g_R^2$       & $0.03076\pm0.00110$ & $0.03006$ & $0.6$ & $0.03034$ & $0.4$ & $0.03037$ & $0.4$ \\
     $g_V^{\nu e}$   & $-0.040\pm0.015$  & $-0.03936$ & $0.0$ & $-0.03804$ & $-0.1$ & $-0.03780$ & $-0.1$ \\
     $g_A^{\nu e}$   & $0.507\pm0.014$   & $-0.5064$ & $0.0$ & $-0.5071$ & $0.0$ & $-0.5071$ & $0.0$ \\
    \hline
\end{tabular}
\label{table_observ} 
}

We scan the parameter space for $M_{Z^\prime}<10$ TeV, and obtain the
allowed deviation at 3$\sigma$ level of the observables including
$M_W$, $\Gamma_Z$, $R_{(e,\mu,\tau,b,c)}$, $\sigma_{had}$,
$A^{FB}_{(e,\mu,\tau,c,b,s)}$, $A_{(e,\mu,\tau,b,c,s)}$, $g_{(L,R)}^2$
and $g_{(V,A)}^{\nu e}$.  In the allowed region, we get the $\chi^2$
values of every points, and then divide the regions into five
sub-regions according to $\chi^2$, they are $\chi^2<32$,
$32<\chi^2<33$, $33<\chi^2<34$, $34<\chi^2<35$ and $\chi^2>35$.  This
is shown in Fig. \ref{fig:chi2_regions}.  When $\tan\beta$ is 0.5 and
1, $M_{Z^\prime}$ can be as low as less than $500$ GeV.  But as
mentioned before, we will concentrate exclusively on the region where
$M_{Z^\prime}$ is larger than 1 TeV.  Since the $\chi^2$ value for the
SM is 32.01, so the blue regions can be regarded as the regions where
our model has a better fit than the SM.  Besides the blue areas, there
are still large regions equally consistent with the current weak scale
experiments as the SM.  The minimal $\chi^2$ values in the allowed
regions for $\tan\beta=0.5, 1, 2, 5, 10, 20, 35, 50$ are 31.82, 31.89,
31.88, 31.87, 31.90, 31.92, 31.92 and 31.92, respectively.  To be
concrete, we present the experimental~\cite{Yao:2006px} and predicted
values of the $Z$-pole observables for the SM~\cite{Yao:2006px} and
our model with $\tan\beta$=2 and 50 for the best fits in
Table~\ref{table_observ}.

\FIGURE{
\centering
\includegraphics[width=0.42\textwidth]{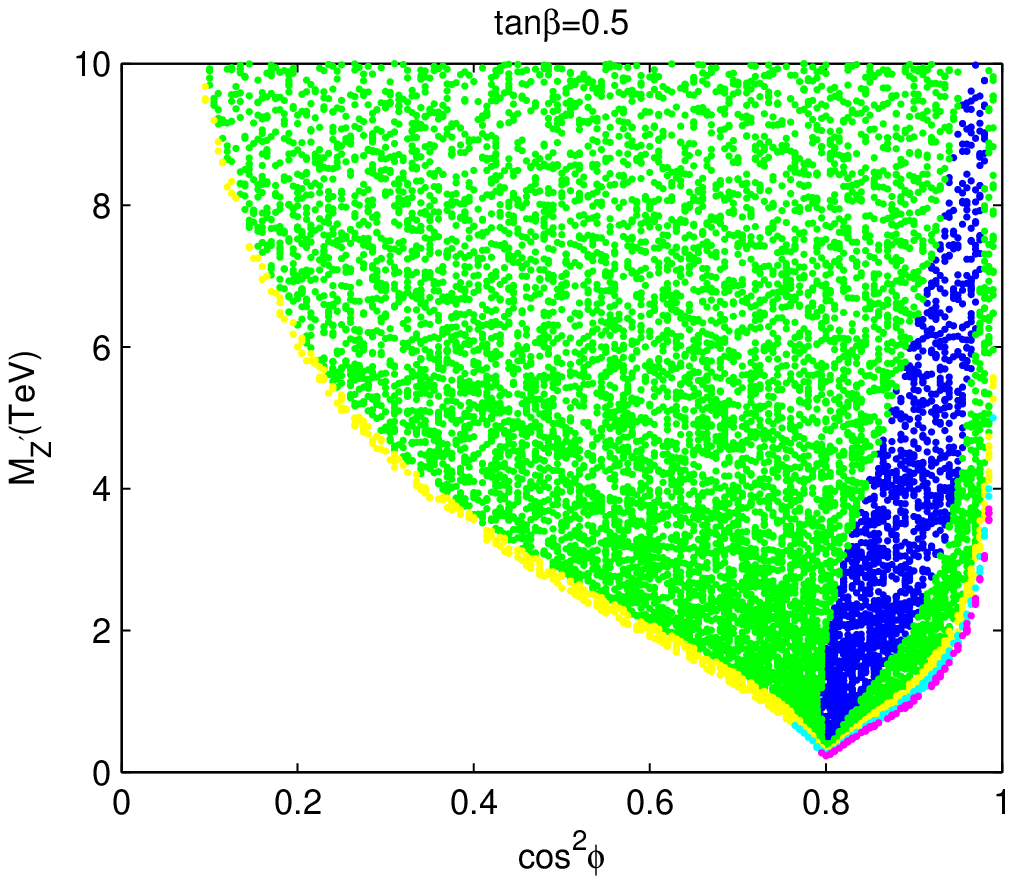}
\includegraphics[width=0.42\textwidth]{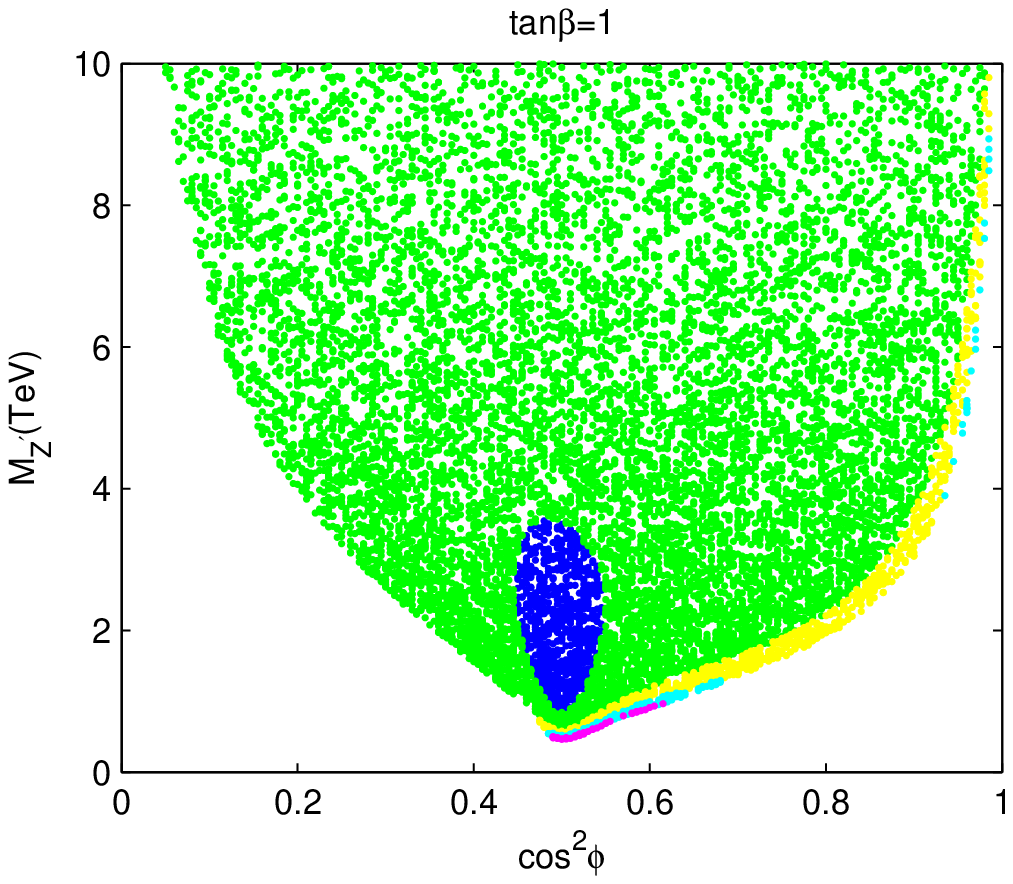}
\includegraphics[width=0.42\textwidth]{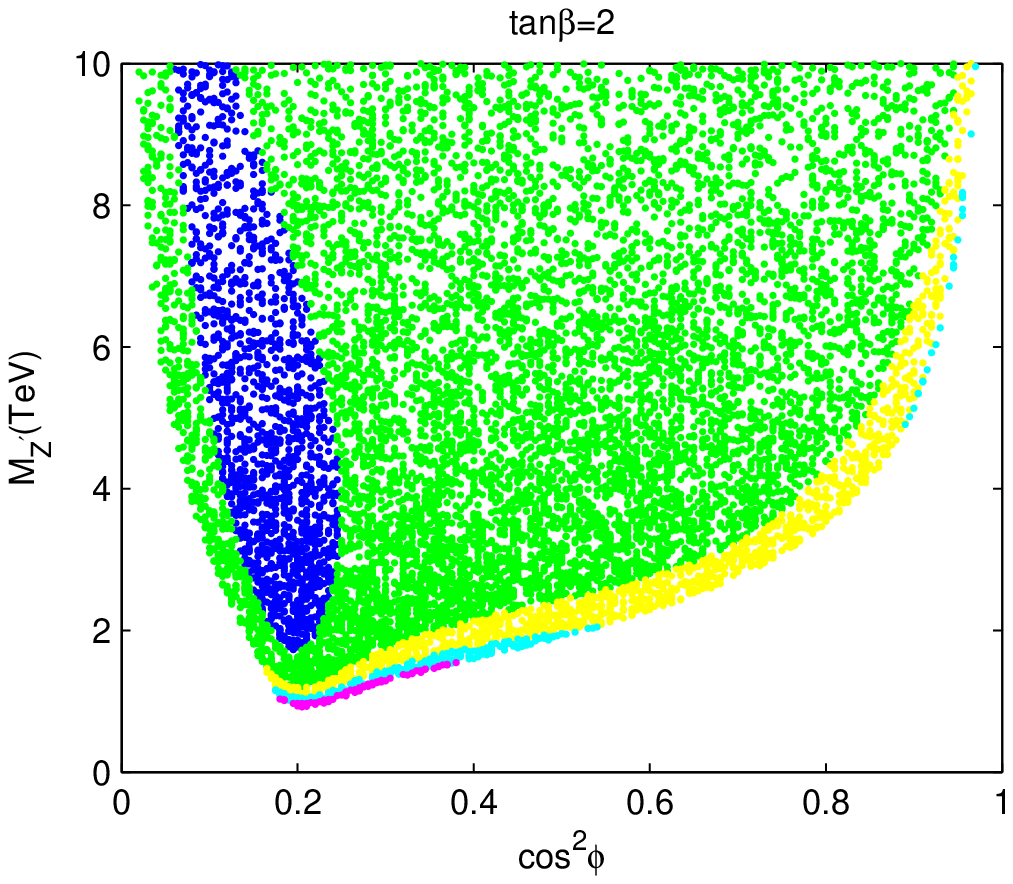}
\includegraphics[width=0.42\textwidth]{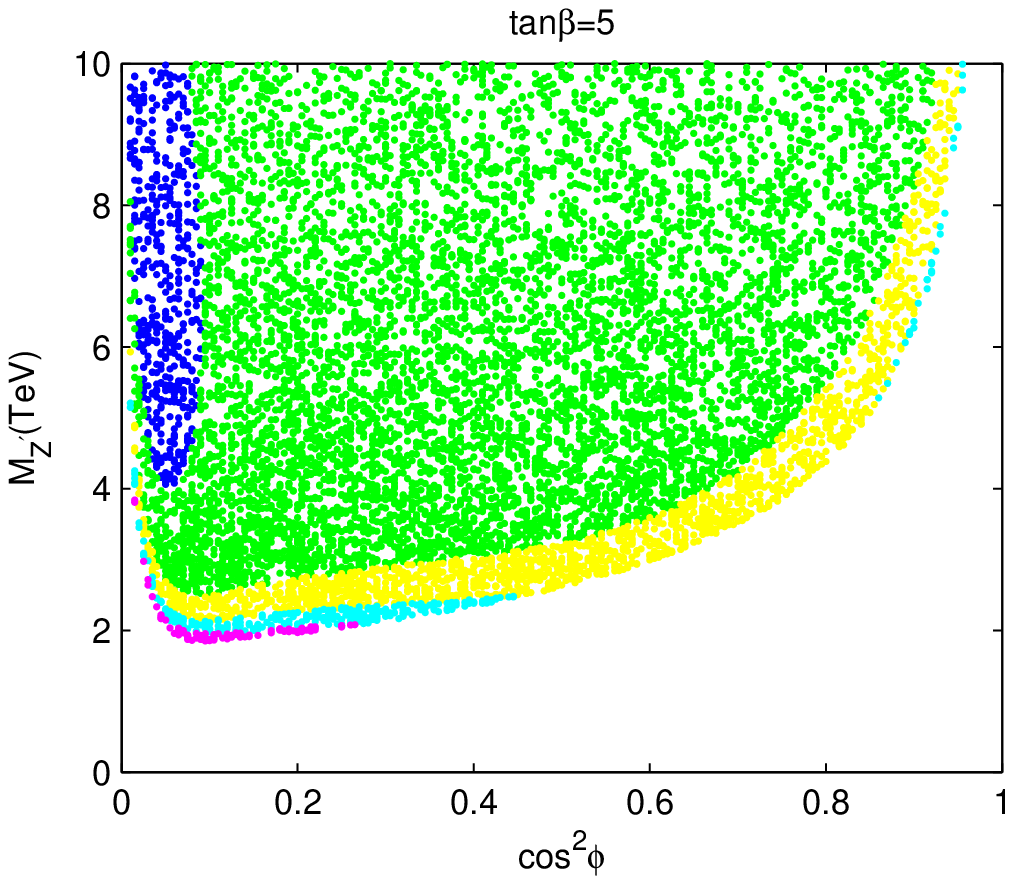}
\includegraphics[width=0.42\textwidth]{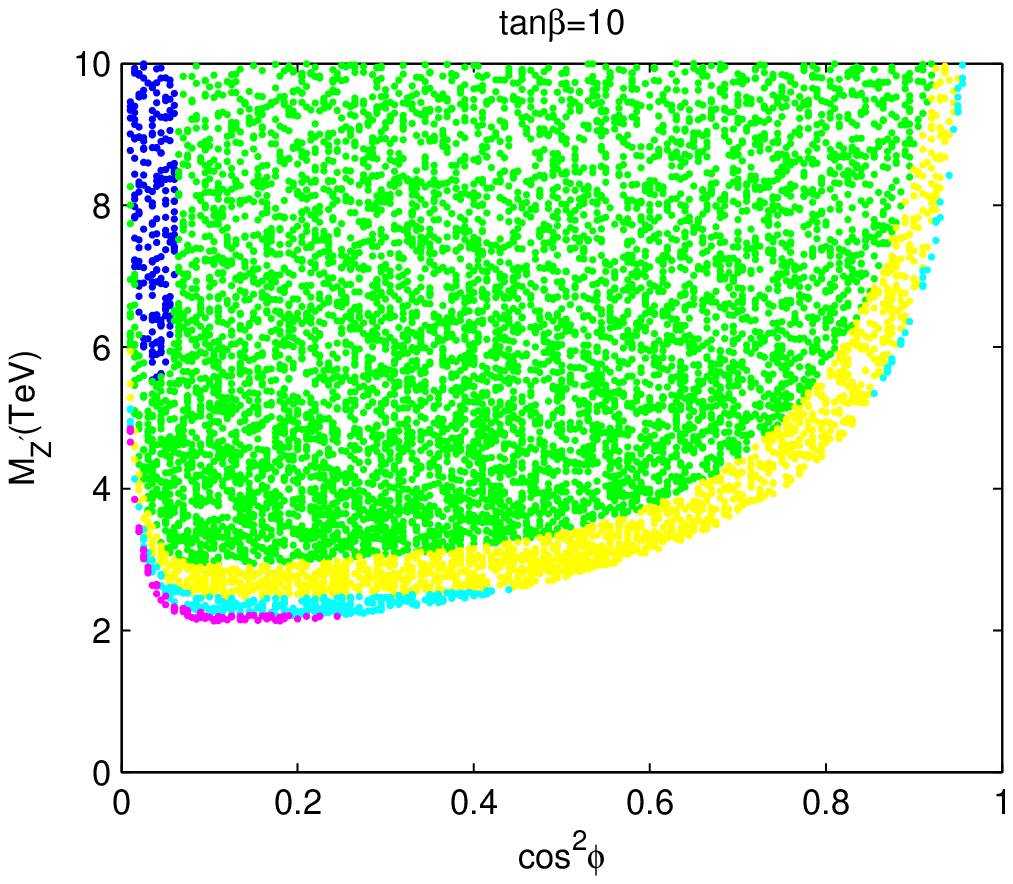}
\includegraphics[width=0.42\textwidth]{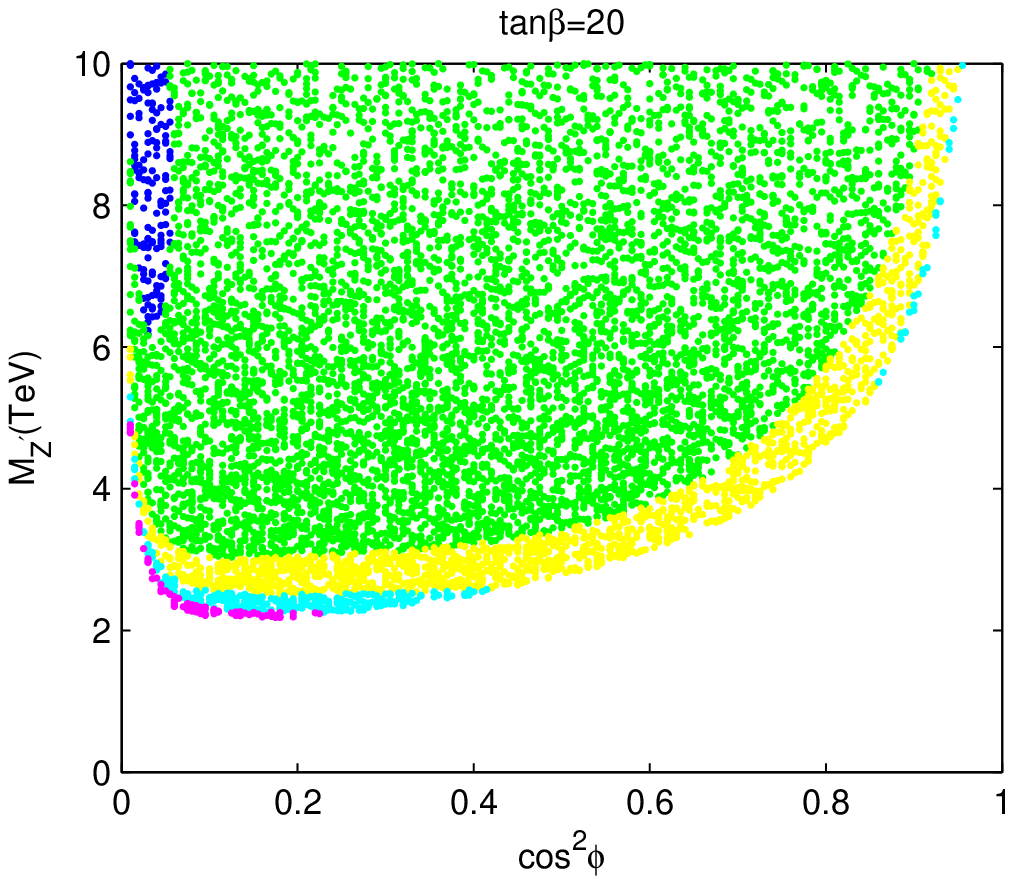}
\includegraphics[width=0.42\textwidth]{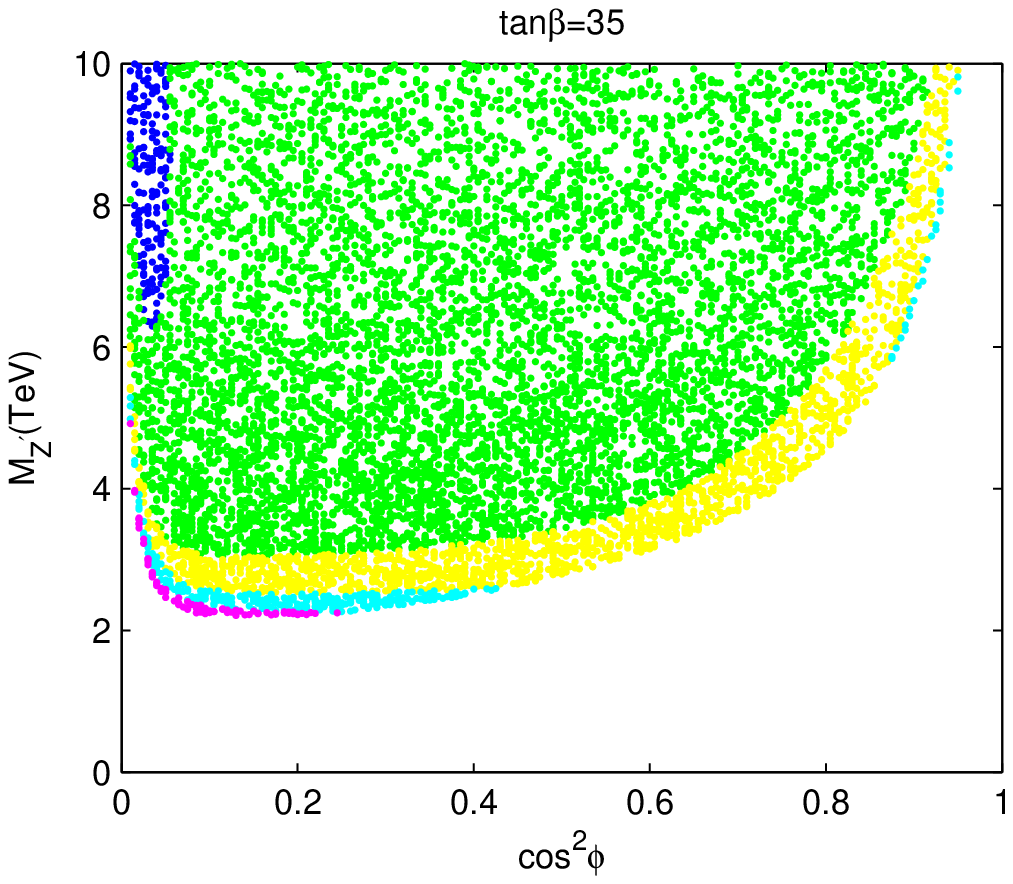}
\includegraphics[width=0.42\textwidth]{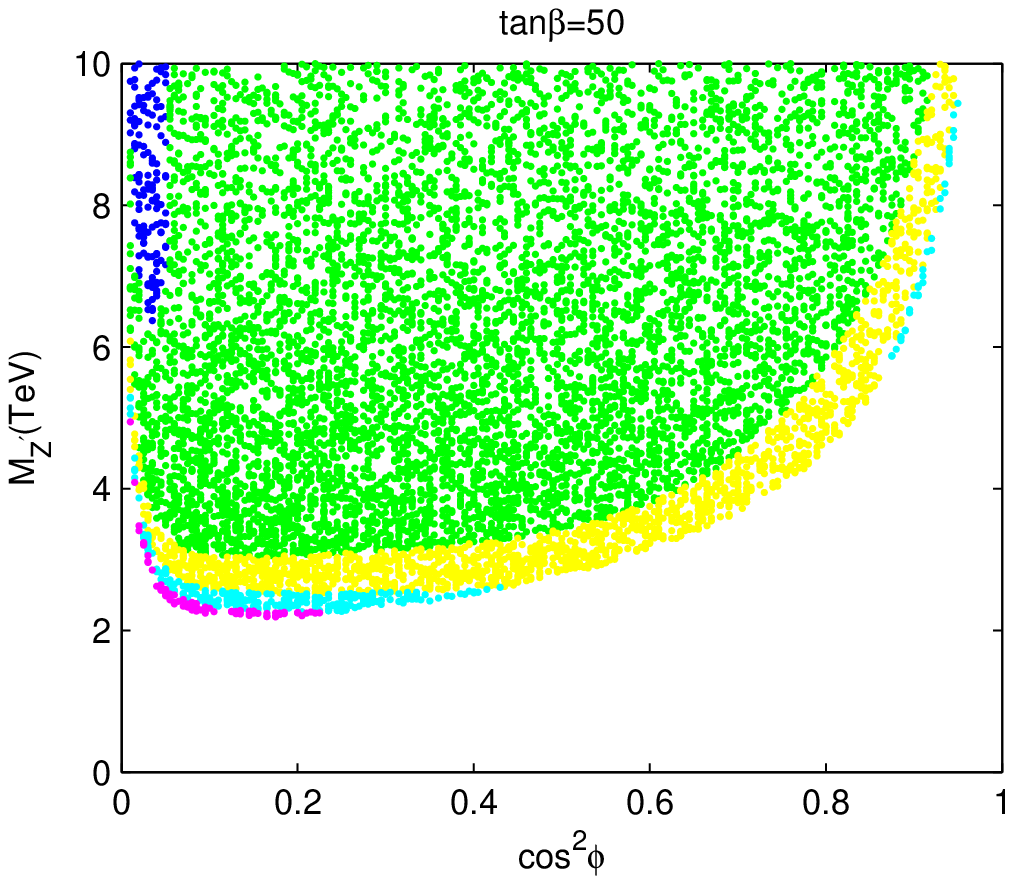}
\caption{Allowed regions of the parameter space for various
  $\tan\beta$.  In all figures, the colors denote regions with
  different $\chi^2$ ranges, blue: $\chi^2<32$, green: $32<\chi^2<33$,
  yellow: $33<\chi^2<34$, cyan: $34<\chi^2<35$, magenta: $\chi^2>35$.}
\label{fig:chi2_regions}
}

The gauge group of the model can be regarded as $SU(3)_C \times
SU(2)_L \times U(1)_Y \times U(1)^\prime$, where the gauge bosons of
$U(1)_Y$ and $U(1)^\prime$ are $B_\mu$ and $\widetilde B_\mu$, as
given in Eq. (\ref{B_definition}).  In general there is a mixing
between $Z$ and $Z^\prime$, and the mixing angle is constrained to be
$10^{-3}$ or smaller \cite{Yao:2006px}.  In our model, this is
achieved when $M_{Z^\prime}$ is larger than $1$ TeV.  Interestingly,
we note that when $\tan\phi = \tan\beta$, there is no mixing between
$Z$ and $Z^\prime$, and then the electroweak precision constraints can
be relaxed as well.

\section{Quark mixing and FCNC \label{sec:FCNC}}

In our model, there are no intrinsic quark Yukawa mixings between the
first two families and the third family. These mixings can be generated
by introducing additional Higgs doublet fields $\Phi_3$, $\Phi_4$,
$\Phi_5$, and $\Phi_6$ with quantum numbers $(1,2,-1/6,-1/3)$, $(1,
2,-1/3,-1/6)$, $(1, 2,-1/6,2/3)$, and $(1, 2,2/3,-1/6)$,
respectively. If we introduce the Higgs doublet fields $\Phi_3$ and
$\Phi_4$, the quark CKM mixings can arise from the down-type quark mass
matrix from the following Yukawa couplings
\begin{eqnarray}
 - {\cal L}_{Yukawa} &=&  Y^d_{3i} \bar {d}_{3R} \Phi_3 Q_{iL}
+ Y^d_{i3} \bar {d}_{iR} \Phi_4  Q_{3L} + h.c. ~.~\,
\end{eqnarray}
Even if we merely introduce $\Phi_3$ or $\Phi_4$, we can still
generate the quark CKM mixings by choosing suitable Yukawa couplings.
Similarly, if we introduce the Higgs doublet fields $\Phi_5$ and
$\Phi_6$, the quark CKM mixings can arise from up-type quark mass
matrix.

Without adding extra Higgs doublet fields, we can also generate the
quark CKM mixings by introducing additional Higgs singlet field and
heavy vector-like fields.  For example, we introduce a SM singlet
Higgs field $S$ with quantum numbers $(1,1,-1/6,1/6)$, and vector-like
fields ($Q_X$, $\overline{Q}_X$), and ($Q'_X$, $\overline{Q}'_X$) with
the following quantum numbers
\begin{eqnarray}
& Q_{X}: (3, 2,~1/6,~0) ~, \hspace{1cm}
\overline{Q}_{X}: (\bar 3, 2,-1/6,~0) ~, \nonumber \\
& {Q}'_{X}: (3, 2,~0,~1/6) ~, \hspace{1cm}
\overline{Q}'_{X}: (\bar 3, 2,~0,-1/6) ~.
\end{eqnarray}
There are new Yukawa coupling terms as the following
\begin{eqnarray}
 - {\cal L}_{Yukawa} &=&
 y_{XQ}^i S \overline{Q}'_{X} Q_i + y_{XQ}^3 S^{\dagger}
  \overline{Q}_{X} Q_3 + y_{Xu}^i \bar {u}_{iR} \Phi_1 Q_{X}
  + y_{Xu}^3 \bar {u}_{3R} \Phi_2 Q_{X}'
  \nonumber \\ &+&
   y_{Xd}^i \bar {d}_{iR} \widetilde \Phi_1 Q_{X}
 + y_{Xd}^3 \bar {d}_{3R} \widetilde \Phi_2 Q_{X}'
+ M_X \overline{Q}_{X} Q_X + M_{X}'  \overline{Q}'_{X} Q'_X + h.c.~,
\end{eqnarray}
where the masses $M_X$ and $M_X'$ can be generated by introducing
another SM singlet Higgs field. For simplicity, we neglect the mixings
between $Q_X$ and $Q_{iL}$, and between $Q_X'$ and $Q_{3L}$.  After we
integrate out the vector-like fields ($Q_X$, $\overline{Q}_X$) and
($Q'_X$, $\overline{Q}'_X$), we obtain
\begin{eqnarray}
 - {\cal L}_{Yukawa} &=& -{1\over {M_X}}\left(y_{Xu}^i y_{XQ}^3
\bar {u}_{iR} \Phi_1 S^{\dagger} Q_3 + y_{Xd}^i y_{XQ}^3 \bar
{d}_{iR} \widetilde \Phi_1 S^{\dagger} Q_3 \right) \nonumber \\ &-&
{1\over {M_X'}}\left(y_{Xu}^3 y_{XQ}^i \bar {u}_{3R} \Phi_2 S Q_i +
y_{Xd}^3 y_{XQ}^i \bar {d}_{3R} \widetilde \Phi_2 S Q_i \right)
 + h.c.~.~\,
\end{eqnarray}
Thus, we generate the SM quark CKM mixings after $S$ obtains a VEV.

The mismatch between the flavor eigenstates and mass eigenstates of
the quarks will result in tree-level flavor-changing $Z'$ couplings
because the U(1)$^\prime$ charges of the third-generation quarks are
different from those of the first two generations.  We can reduce
uncertainties from the up sector by assuming no difference between the
flavor and mass eigenstates for the up-type quarks, corresponding to
the case where only the $\Phi_3$ and $\Phi_4$ Higgs fields are
introduced, as described above.  See, however,
Ref.~\cite{Arhrib:2006sg} for the situation where the up sector mixings
are identical to the CKM mixings, and its consequences to single-top
production.  In this case, the converting matrix for the down-type
quarks is simply the CKM matrix.  Following the discussions and
notations in Refs.~\cite{Barger:2003hg,Cheung:2006tm}, the dominant
off-diagonal $Z'$ coupling is between the left-handed bottom and
strange quarks due to the hierarchical structure in the CKM matrix
\begin{eqnarray}
B_L^{sb} = \delta_L Q_{d_L} V_{tb} V_{ts}^* ~,
\end{eqnarray}
where the combination
\begin{eqnarray}
\delta_L Q_{d_L} =
\frac{e}{3 \sin2\phi \cos\theta} ~.
\end{eqnarray}
It is interesting to note that there is no $\tan\beta$ dependence in
the FCNC coupling $B_L^{sb}$.  Such a coupling can contribute to
processes involving $b \to s$ transitions.  In view of their
precisions and less theoretical uncertainties, the latest experimental
data of the $B_s$-$\overline{B}_s$ oscillation \cite{Abulencia:2006ze}
and $B \to X_s \ell^+ \ell^-$ decay
\cite{Aubert:2004it,Iwasaki:2005sy} are used here to constrain the
parameters $\cos\phi$ and $M_{Z'}$ appearing in the above expressions.

Given the value $\Delta M_s^{\rm exp} = 17.77 \pm 0.10 \pm 0.07$
ps$^{-1}$ reported by the CDF Collaboration \cite{Abulencia:2006ze}
and the SM expectation $\Delta M_s^{\rm SM} = 19.52 \pm 5.28$
ps$^{-1}$, we determine the ratio $\Delta M_s^{\rm exp} / \Delta
M_s^{\rm SM} = 0.89 \pm 0.24$.  Suppose there are only left-handed
flavor-changing $Z'$ couplings in our model, we find that the allowed
parameter space is above the solid curve in Fig.~\ref{fig:chi2lh}.
Furthermore, if we assume no mixing in the lepton sector, the
constraint given by the $B \to X_s \ell^+ \ell^-$ decays renders the
space above the dashed green curve.  In this case, the $\Delta M_s$
constraint is stronger for $Z'$ mass above about 500 GeV.

\FIGURE{
\centering
\includegraphics[width=10cm]{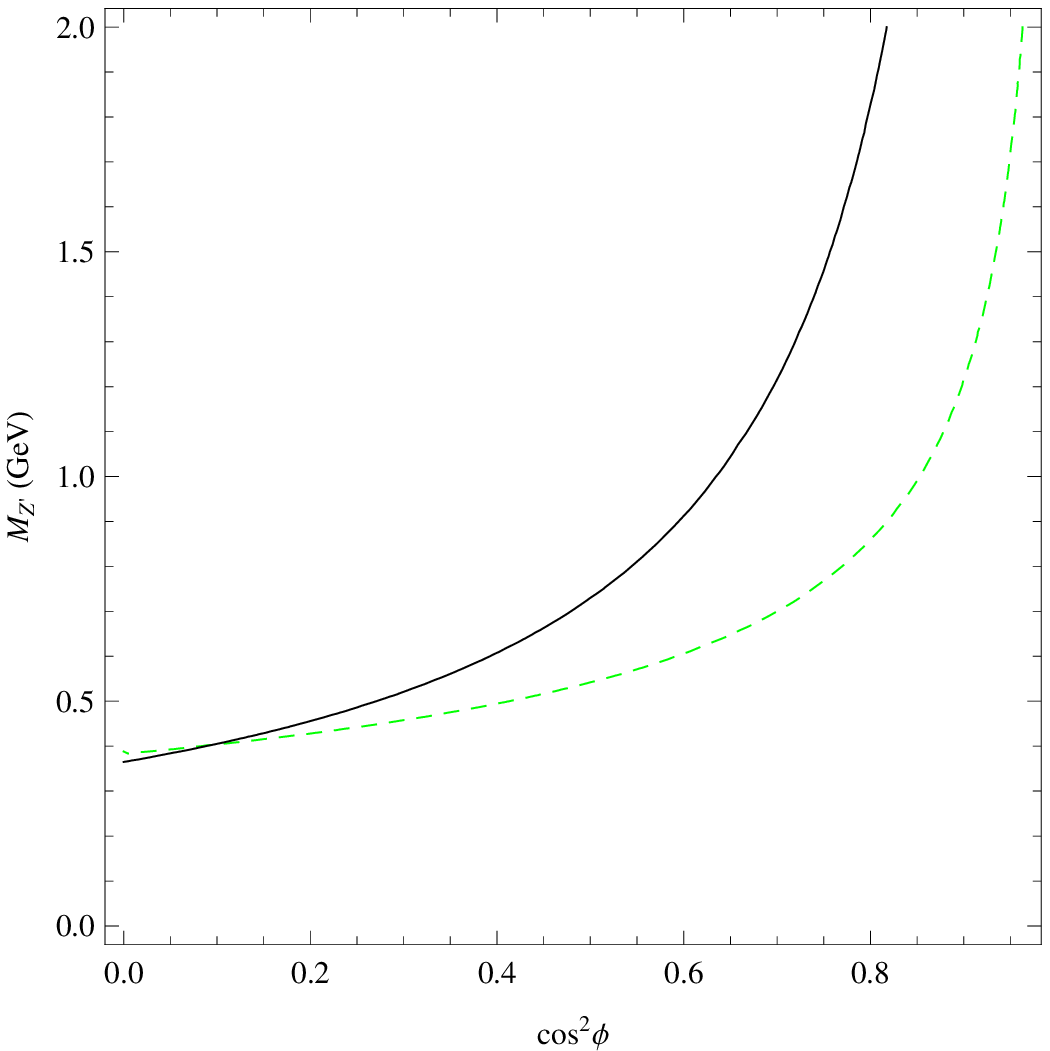}
\caption{Constraints on the model parameters $(\cos\phi,M_{Z'})$ using
  the $B_s$-$\overline{B}_s$ oscillation (solid black curve) and the
  averaged $B \to X_s \ell^+ \ell^-$ decay rate (dashed green curve).
  Here the model only contains the left-handed FCNC couplings for the down
  quark sector.  The allowed parameter space is above the curves.}
\label{fig:chi2lh}
}

If there is also a mismatch between the flavor and mass eigenstates
for the right-handed quarks, then there will be flavor-changing $Z'$
couplings for them and thus additional contributions to the
above-mentioned processes.  However, such an analysis cannot be done
without knowing the mixing information.  As an example, if we assume
the same CKM matrix for the right-handed quarks, the $Z'$-$s_R$-$b_R$
coupling is
\begin{eqnarray}
B_R^{sb} = \delta_R Q_{d_R} V_{tb} V_{ts}^* ~,
\end{eqnarray}
where the combination
\begin{eqnarray}
\delta_R Q_{d_R} =
-\frac{2e}{3 \sin2\phi \cos\theta} ~.
\end{eqnarray}
$B_R^{sb}$ is also $\tan\beta$ independent.  With the inclusion of
such flavor-changing $Z'$ couplings to both left-handed and
right-handed down-type quarks, the allowed parameter space on the
$\cos\phi$-$M_{Z'}$ plane is given in Fig.~\ref{fig:chi2lrh}.

The $B_s$-$\overline{B}_s$ oscillation data exclude the region below
the solid blue curve and the region between the dash-dotted red
curves.  This slightly more complicated condition results from the
introduction of new operators $[\overline{s} \gamma^\mu (1-\gamma^5)
b] [\overline{s} \gamma_\mu (1+\gamma^5) b]$, $[\overline{s}
(1-\gamma^5) b] [\overline{s} (1+\gamma^5) b]$, and $[\overline{s}
\gamma^\mu (1+\gamma^5) b] [\overline{s} \gamma_\mu (1+\gamma^5) b]$
in the model.  The first two new operators have destructive
interference with the third new operator and the standard model
operator $[\overline{s} \gamma^\mu (1-\gamma^5) b] [\overline{s}
\gamma_\mu (1-\gamma^5) b]$.  Moreover, the renormalization group
running enhances the Wilson coefficient of the former.  The $B \to X_s
\ell^+ \ell^-$ decays in this case has a stronger constraint than the
$B_s$ mixing when $\cos\phi < 0.6$.

\FIGURE{
\centering
\includegraphics[width=10cm]{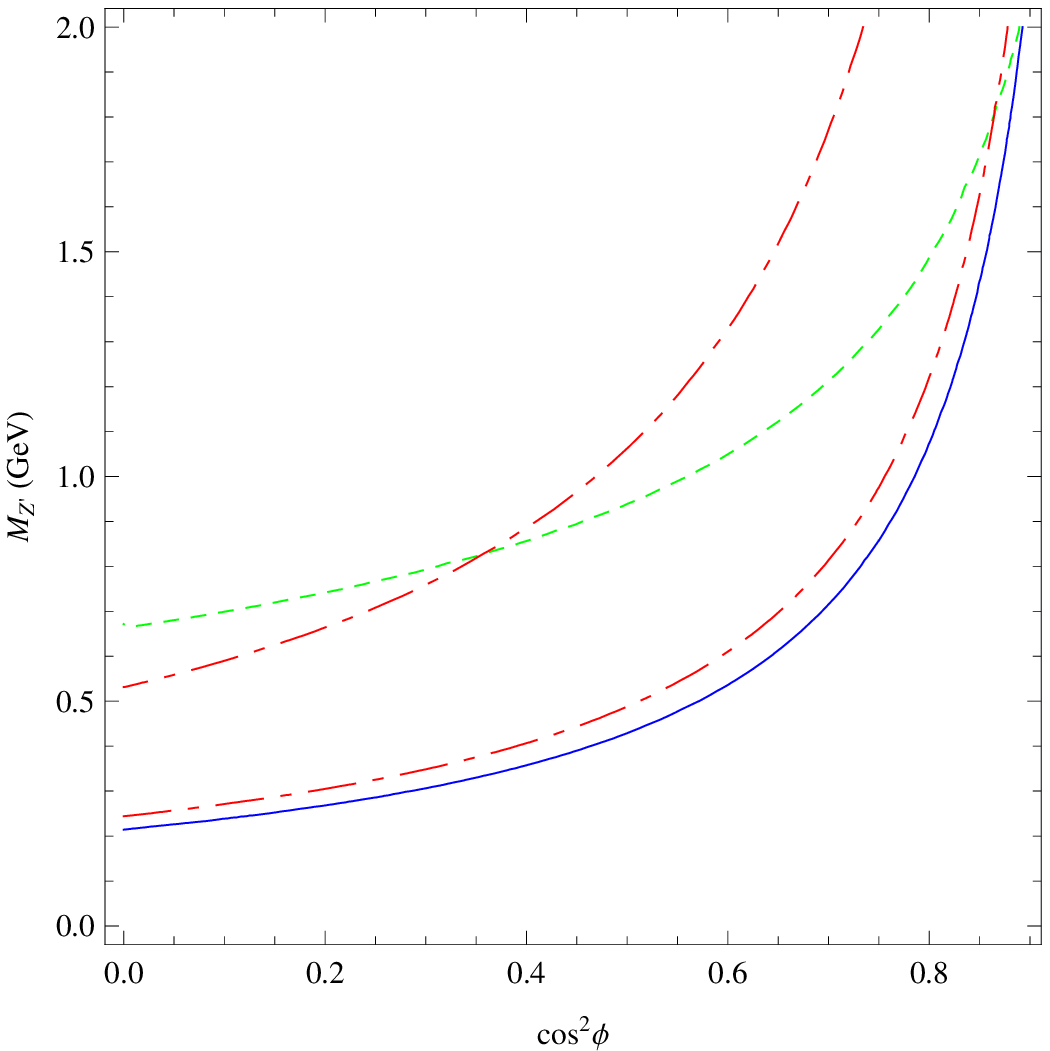}
\caption{Constraints on the model parameters $(\cos\phi,M_{Z'})$ using
  the $B_s$-$\overline{B}_s$ oscillation (solid blue and dash-dotted
  red curves) and the averaged $B \to X_s \ell^+ \ell^-$ decay rate
  (dashed green curve).  Here the model contains both the left-handed and
  right-handed FCNC couplings for the down quark sector.  The allowed
  parameter space is above the solid blue and dashed green curves with
  the region between the dash-dotted red curves being excluded.}
\label{fig:chi2lrh}
}

\section{$Z^\prime$ production \label{sec:prod}}

As the previous sections indicate, for a specific choice of
$\cos\phi$, $\epsilon_2$ is allowed to vary within a certain range
while still producing a reasonable fit to the experimental
observables.  When we consider the $Z^\prime$ production at the LHC,
we present our results for four values of $\cos\phi$, 0.2, 0.4, 0.6
and 0.8.  For each value of $\cos\phi$, $\epsilon_2$ is adjusted to
generate the corresponding $M_{Z^\prime}$ to be within the range of 1
TeV to 5 TeV.  Similar to other types of $Z^\prime$
models~\cite{Rizzo:2006nw}, the $Z^\prime$ in the top hypercharge
model can be searched for through the Drell-Yan processes.  For the
mass range we consider, the decay width of the $Z^\prime$ is typically
a few hundred GeV.  We apply the simple cuts of requiring the
transverse momenta of the outgoing leptons to be larger than 20 GeV,
the absolute value of the rapidities of the leptons to be less than
2.5, and the invariant mass of the lepton pair to be between
$M_{Z^\prime} - 1/2 \Gamma_{Z^\prime}$ and $M_{Z^\prime} + 1/2
\Gamma_{Z^\prime}$.  After these cuts, the SM background becomes
negligible compared to the signal.

We calculate the $p p \to e^+ e^-$ and $\mu^- \mu^+$ cross sections
and deduce the significance for discovery based on the statistical
errors.  In Fig.~\ref{fig:5sig}, we show the total luminosities
required for a $5\sigma$ discovery of different $\cos\phi$ and
$M_{Z^\prime}$ in the top hypercharge model.  With 100 fb$^{-1}$, the
discovery reaches for $\cos\phi = 0.8$, 0.6, 0.4 and 0.2 are about
4.1, 5.0, 5.8 and 6.9 TeV, respectively.  The production cross
sections have a strong dependence on $\cos\phi$, as can be seen in
Eq.~(\ref{co_derivative_mass}) that the $Z^\prime$ coupling to the
first two generations of quarks are proportional to $\tan\phi$.
Therefore, small $\cos\phi$ will enhance this coupling and large cross
sections follow.

\FIGURE{
\centering
\includegraphics[width=10cm]{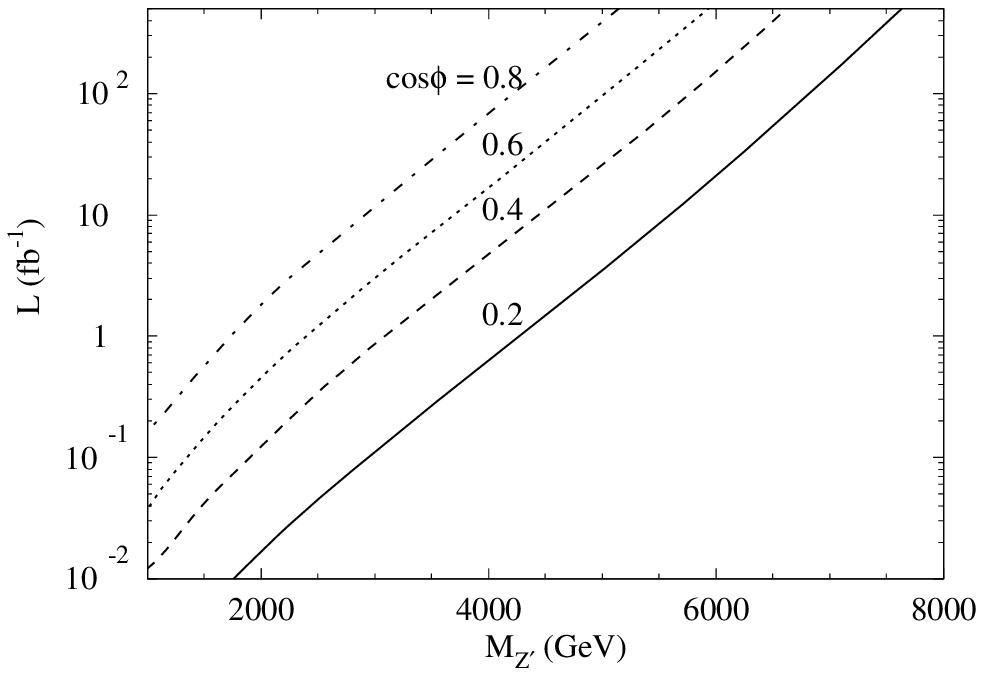}
\caption{The total luminosity required for a 5$\sigma$ discovery
  through the Drell-Yan process as a function of the $Z^\prime$ mass
  for $\cos\phi = 0.2, 0.4, 0.6$, and $0.8$.}
\label{fig:5sig}
}

\section{Dark matter and Higgs masses \label{sec:DM-Higgs}}

Finally, we comment on the issues of dark matter candidate and Higgs
boson masses in this model.  As to be seen below, this part of
discussions involves more independent parameters.  Detailed numerical
analyses are beyond the scope of this work and will be presented
elsewhere \cite{TBD}.

The dark matter candidate is a particle whose stability can be
achieved using an additional discrete symmetry.  However, if the
discrete symmetry is global but not gauged, this symmetry can be
violated by non-renormalizable higher-dimensional operators due to
quantum gravity effects.  In our model, we can introduce a SM singlet
scalar field $\phi$ with quantum number $(1,1,-1/4,1/4)$.  The
relevant Lagrangian for $\phi$ and its couplings to $\Sigma$ is
\begin{eqnarray}
 - {\cal L}_{Yukawa} &=& m_{\phi}^2 |\phi|^2 +
 {{\lambda_{\phi}}\over 2}
 |\phi|^4 + \lambda'_{\phi} |\phi|^2 |\Sigma|^2 +
 \left({\widetilde m}_{\phi} \phi^2 \Sigma + h. c.\right)~.~\,
\end{eqnarray}
Thus, after the $U(1)_1\times U(1)_2$ gauge symmetry is broken down to
$U(1)_Y$, $\phi$ becomes a stable particle due to the gauged $Z_2$
symmetry under which $\phi$ goes to $-\phi$ and the other fields
are invariant.

Similar to the singlet-field dark matter~\cite{McDonald:1993ex,
  Burgess:2000yq, Davoudiasl:2004be}, $\phi$ can possibly give the
correct dark matter density, which deserves further study in details
because of the additional $U(1)$ gauge symmetry.

Aside from the above-mentioned $Z_2$-odd scalar field, there are at
least three Higgs fields in the model, namely $\Sigma$, $\Phi_1$ and
$\Phi_2$.  The most general renormalizable Higgs potential is
\begin{eqnarray}
V_{scalar} &=& -m_{11}^2 \Phi_1^\dagger \Phi_1 - m_{22}^2
\Phi_2^\dagger \Phi_2 - \left[A_m \Phi_1^\dagger \Sigma \Phi_2 +
  h.c.\right] 
\nonumber \\
&+& {1 \over 2} \lambda_1 \left(\Phi_1^\dagger \Phi_1\right)^2 + {1
\over 2} \lambda_2 \left(\Phi_2^\dagger \Phi_2\right)^2 + \lambda_3
\left(\Phi_1^\dagger \Phi_1\right)\left(\Phi_2^\dagger \Phi_2\right)
+ \lambda_4 \left(\Phi_1^\dagger \Phi_2\right)\left(\Phi_2^\dagger
\Phi_1\right) \nonumber \\
&+& \lambda_5 \left(\Sigma^\dagger \Sigma\right)\left(\Phi_1^\dagger
\Phi_1\right) + \lambda_6 \left(\Sigma^\dagger
\Sigma\right)\left(\Phi_2^\dagger \Phi_2\right) - m_\Sigma^2
\Sigma^\dagger \Sigma + {1\over2}\lambda_\Sigma
\left(\Sigma^\dagger\Sigma\right)^2 ~,
\end{eqnarray}
where $A_m$ has mass dimension one, and the coefficients $\lambda_i$
are real dimensionless quantities.  We ignore the possibility of
explicit CP-violating effects in the Higgs potential by choosing $A_m$
to be real, although in general it can be complex.  At the minimum of
the potential, we expand the Higgs fields as follows
\begin{eqnarray}
\Sigma = {1\over\sqrt{2}}\left(u+\sigma+i\xi\right)~,~\,
\end{eqnarray}
\begin{eqnarray}
\Phi_1=\left( \begin{array}{c} \phi_1^+ \\
{\left( v_1+\phi_1+i\varphi_1 \right) / \sqrt{2}} \end{array}
\right)~,~~
\Phi_2=\left( \begin{array}{c} \phi_2^+ \\
\left(v_2+\phi_2+i\varphi_2\right)/\sqrt{2} \end{array} \right) ~.
\end{eqnarray}
There are ten degrees of freedom in the three Higgs fields.  After the
$SU(2)_L\times U(1)_1 \times U(1)_2$ gauge symmetry is broken down to
$U(1)_{EM}$, there will be left with six physical Higgs fields: three
CP-even ones ($H_1$, $H_2$ and $H_3$), a CP-odd one ($A$), and a
charged pair ($H^\pm$).  The mass-squared matrix of the CP-even Higgs
fields in the basis of $(\sigma,\phi_1,\phi_2)$ is
\begin{eqnarray}
{\cal M}_{CP-even}^2 = \left( \begin{array}{ccc} {A_m v_1 v_2 \over
\sqrt{2} u} + \lambda_\Sigma u^2 & -{A_m v_2 \over \sqrt{2}} +
\lambda_5 u v_1 & -{A_m v_1 \over \sqrt{2}} + \lambda_6 u v_2 \\
-{A_m v_2 \over \sqrt{2}} + \lambda_5 u v_1 & {A_m u v_2 \over
\sqrt{2} v1} + \lambda_1 v_1^2 & -{A_m u \over
\sqrt{2}}+\left(\lambda_3+\lambda_4\right)v_1 v_2 \\
-{A_m v_1 \over \sqrt{2}} + \lambda_6 u v_2 &  -{A_m u \over
\sqrt{2}}+\left(\lambda_3+\lambda_4\right)v_1 v_2 & {A_m u v_1 \over
\sqrt{2} v_2} + \lambda_2 v_2^2 \end{array} \right) ~.
\end{eqnarray}
The mass-squared matrix of CP-odd Higgs fields in the basis of
$(\xi,\varphi_1,\varphi_2)$ is
\begin{eqnarray}
{\cal M}_{CP-odd}^2 = \frac{A_m}{\sqrt{2}} 
\left( \begin{array}{ccc} {v_1 v_2  u} & -{ v_2 } & {v_1 }
\\
-{v_2 } & {u v_2 v_1} & -{ u
} \\
{v_1 } & -{u } & { u v_1
 v_2}
\end{array}
\right) ~.
\end{eqnarray}
Thus, there are two massless Goldstone bosons and one CP-odd Higgs
field $A$ with the following mass
\begin{eqnarray}
m_A^2 &=& {A_m \left( v_1^2 v_2^2 + u^2 v_1^2 + u^2 v_2^2 \right)
\over \sqrt{2} u v_1 v_2} ~.
\end{eqnarray}
The mass-squared matrix of charged Higgs fields in the basis of
$(\phi_1^\pm,~\phi_2^\pm)$ is
\begin{eqnarray}
{\cal M}_{charged}^2 = \left( \begin{array}{cc}  {A_m v_2 u \over
\sqrt{2} v_1} - {1 \over 2}\lambda_4 v_2^2 & -{A_m u \over
\sqrt{2}}+{1\over 2} \lambda_4 v_1 v_2 \\
-{A_m u \over \sqrt{2}}+{1\over 2} \lambda_4 v_1 v_2 & {A_m v_1 u
\over \sqrt{2} v_2} - {1 \over 2}\lambda_4 v_1^2
\end{array} \right) ~.~\,
\end{eqnarray}
So, we have one pair of massless charged Goldstone bosons, and one
pair of charged Higgs bosons with mass
\begin{eqnarray}
m_{H^\pm}^2 = {A_m u (v_1^2+v_2^2)\over \sqrt{2} v_1 v_2} - {1\over
2}\lambda_4(v_1^2+v_2^2) ~.~\,
\end{eqnarray}

\section{Conclusions \label{sec:summary}}

In this paper, we proposed a top hypercharge model with the gauge
symmetry $SU(3)_C \times SU(2)_L \times U(1)_1 \times U(1)_2$.  The
first two families of the SM fermions are charged under $U(1)_1$ while
the third family is charged under $U(1)_2$.  We break the $U(1)_1
\times U(1)_2$ gauge symmetry down to the $U(1)_Y$ gauge symmetry by
giving a VEV to the Higgs field $\Sigma$, which is an $SU(2)_L$
singlet but charged under both $U(1)$ groups.  We performed a global
fit to the electroweak observables at the $Z$-pole in this model and
found that in some regions of the parameter space
(Fig.~\ref{fig:chi2_regions}) the top hypercharge model provides a
better fit than the SM.  In view of the possibility of a large FCNC
associated with the $b \to s$ transition, we studied the constraints
by considering the $B_s$ mixing and $B \to X_s \ell^+ \ell^-$ decays.
They generally excluded the regions with $m_{Z^\prime} < 500$ GeV and
large $\cos^2\phi$.  The production and detection of such a $Z^\prime$
boson through the Drell-Yan process at the LHC was calculated for $0.2
\le \cos\phi \le 0.8$.  With an integrated luminosity of $100$
fb$^{-1}$, the $5 \sigma$ discovery reach ranged from $4.1$ to $6.9$
TeV.  We also commented on the possibility that a scalar field charged
odd under a gauged $Z_2$ symmetry derived from the extra $U(1)$
symmetry may serve as a DM candidate, and the tree-level mass spectrum
of the Higgs boson fields.  These issues deserve further
investigations in the future.

\section*{Acknowledgments}

We would like to thank Zhao-Feng Kang very much for collaborations in
the early stage of this project.  This research was supported in part
by the National Science Council of Taiwan, R.O.C. under Grant No.~NSC
95-2112-M-008-008, by the U.S.~Department of Energy under Grants
No.~DE-FG02-96ER40969, and by the Cambridge-Mitchell Collaboration in
Theoretical Cosmology.  C.-W.~C. would like to thank the hospitality
of the Fermilab Theory Group, U.S.A. and National Center for Theoretical
Sciences, Hsinchu, Taiwan where part of this work is carried out.

\end{document}